\documentclass[journal]{IEEEtran}
\usepackage[utf8]{inputenc} 
\usepackage[T1]{fontenc}    
\usepackage{hyperref}       
\usepackage{url}            
\usepackage{booktabs}       
\usepackage{amsfonts}       
\usepackage{nicefrac}       
\usepackage{microtype}      
\usepackage{xcolor}    
\usepackage{soul}
\usepackage{subfig}
\usepackage{graphicx}
\usepackage{float}
\usepackage{placeins}
\usepackage{multicol}
\usepackage{multirow}
\usepackage{cite}
\usepackage{array}\usepackage{makecell}
\usepackage{amsmath}
\usepackage{comment}
\usepackage{threeparttable}
\usepackage{tabularx}
\usepackage{setspace}
\usepackage{subfig} 
\usepackage{hyperref}
\usepackage{pgfplots}
\usepackage{colortbl}
\usepackage{algorithm} 
\usepackage{algpseudocode}
\usepackage{graphicx}
\usepackage{adjustbox}
\usepackage{amssymb}
\usepackage{mathtools}
\usepackage{bbm}

\pgfplotsset{compat=1.17}

\usepackage{tikz}
\usetikzlibrary{positioning}
\usetikzlibrary{shapes.geometric, decorations.pathreplacing} 
\tikzset{>=latex}

\usepackage{siunitx}
\usepackage[outline]{contour} 
\contourlength{1.2pt}
\usetikzlibrary{3d} 
\usetikzlibrary{shadows.blur}
\usetikzlibrary{fadings}
\usetikzlibrary{decorations.text} 
\usetikzlibrary{shapes.misc}
\usetikzlibrary{pgfplots.statistics}
\usetikzlibrary{positioning}
\usetikzlibrary{shapes.geometric, decorations.pathreplacing}
\usetikzlibrary{arrows}
\usetikzlibrary{fit}
\usetikzlibrary{calc}

\tikzset{>=latex}
\tikzstyle{xlab}=[below=-1,scale=0.85]
\tikzstyle{ylab}=[left=-1,scale=0.85]

\colorlet{mydarkblue}{blue!40!black}
\colorlet{mylightblue}{mydarkblue!12} 
\colorlet{myred}{red!80!black}
\colorlet{mydarkred}{red!50!black}
\colorlet{mylightred}{mydarkred!12}
\colorlet{mydarkgreen}{green!30!black}
\colorlet{mylightgreen}{mydarkgreen!12}
\colorlet{myorange}{orange!63!black}
\colorlet{mylightorange}{orange!80!black!12}

\usetikzlibrary{patterns} 
\def\hatchsize{4pt}
\makeatletter
\pgfdeclarepatternformonly{myhatch}{%
  \pgfqpoint{-1pt}{-1pt}}{\pgfqpoint{\hatchsize}{\hatchsize}}{\pgfqpoint{\hatchsize}{\hatchsize}
}{%
  \pgfsetlinewidth{0.3pt}
  \pgfpathmoveto{\pgfqpoint{0pt}{\hatchsize}}
  \pgfpathlineto{\pgfqpoint{\hatchsize}{0pt}}
  \pgfusepath{stroke}
}
\makeatother

\tikzfading[
  name=fade out,
  inner color=transparent!0,
  outer color=transparent!100
]
\tikzfading[
  name=fade right,
  left color=transparent!0,
  right color=transparent!100
]

\usetikzlibrary{shadings}
\makeatletter
\pgfdeclareradialshading[tikz@radial@inner,tikz@radial@outer]%
  {myradial}{\pgfpointorigin}{
    color(0bp)=(pgftransparent!0);%
    color(20bp)=(pgftransparent!90);%
    color(30bp)=(pgftransparent!93);%
    color(40bp)=(pgftransparent!97);%
    color(50bp)=(pgftransparent!100)%
  }
\makeatother
\pgfdeclarefading{myradial}{\pgfuseshading{myradial}}%

\def\minwidth{5cm}

\tikzstyle{pooling} = [draw, rectangle, minimum width=\minwidth, minimum height=.75cm, fill=red!20, text centered, anchor=north]

\tikzstyle{activation} = [draw, rectangle, minimum width=\minwidth, minimum height=.75cm, fill=gray!20, text centered, anchor=north]

\tikzstyle{container} = [draw, rectangle, minimum width=\minwidth, minimum height=.75cm, fill=yellow!20, text centered, anchor=north]

\tikzstyle{norm} = [draw, rectangle, minimum width=\minwidth, minimum height=.75cm, fill=blue!20, text centered, anchor=north]

\tikzstyle{conv} = [draw, rectangle, minimum width=\minwidth, minimum height=.75cm, fill=purple!20, text centered, anchor=north]

\tikzstyle{mul} = [draw, rectangle, minimum width=\minwidth, minimum height=.75cm, fill=green!20, text centered, anchor=north]

\tikzstyle{downsample} = [draw, trapezium, trapezium left angle=120, trapezium right angle=120, minimum width = \minwidth, minimum height = .5cm, fill = orange!20, text centered, anchor =north]

\tikzstyle{upsample} = [draw, trapezium, trapezium left angle= 60, trapezium right angle=60, minimum width = \minwidth, minimum height = .5cm, fill = orange!20, text centered, anchor =north]

\tikzstyle{linear} = [draw, rectangle, minimum width=\minwidth, minimum height=.75cm, fill=orange!20, text centered, anchor=north]

\newcommand{\appropto}{\mathrel{\vcenter{
     \offinterlineskip\halign{\hfil$##$\cr
     \propto\cr\noalign{\kern2pt}\sim\cr\noalign{\kern-2pt}}}}}
     
\DeclareCaptionLabelSeparator{newline}{\\}
\makeatletter
\def\maketag@@@#1{\hbox{\m@th\normalfont\footnotesize#1}}
\makeatother

\title{A Novel Contrastive Loss for Zero-Day Network Intrusion Detection}

\author{Jack Wilkie, Hanan Hindy, Craig Michie, Christos Tachtatzis, James Irvine, Robert Atkinson 
\thanks{Jack Wilkie, Craig Michie, Christos Tachtatzis, James Irvine and Robert Atkinson are with the Department of Electronics and Electrical Engineering, University of Strathclyde, Glasgow, United Kingdom.}
\thanks{Hanan Hindy is with Faculty of Computer and Information Sciences, Ain Shams University, Egypt.}
}

\begin{document}
\IEEEoverridecommandlockouts

\IEEEpubid{\begin{minipage}{\textwidth}\ \vspace{1.4cm}\\[10pt]
\centering
\copyright~2026 IEEE. Personal use of this material is permitted. 
Permission from IEEE must be obtained for all other uses, in any current or future media, 
including reprinting/republishing this material for advertising or promotional purposes, 
creating new collective works, for resale or redistribution to servers or lists, 
or reuse of any copyrighted component of this work in other works.
DOI: \url{10.1109/TNSM.2026.3652529}
Link to IEEE Xplore: \url{https://ieeexplore.ieee.org/document/11340750}
\end{minipage}}

\maketitle

\begin{abstract}
     Machine learning has achieved state-of-the-art results in network intrusion detection; however, its performance significantly degrades when confronted by a new attack class--- a zero-day attack. In simple terms, classical machine learning-based approaches are adept at identifying attack classes on which they have been previously trained, but struggle with those not included in their training data. One approach to addressing this shortcoming is to utilise anomaly detectors which train exclusively on benign data with the goal of generalising to all attack classes--- both known and zero-day. However, this comes at the expense of a prohibitively high false positive rate. This work proposes a novel contrastive loss function which is able to maintain the advantages of other contrastive learning-based approaches (robustness to imbalanced data) but can also generalise to zero-day attacks. Unlike anomaly detectors, this model learns the distributions of benign traffic using both benign and known malign samples, i.e. other well-known attack classes (not including the zero-day class), and consequently, achieves significant performance improvements. The proposed approach is experimentally verified on the Lycos2017 dataset where it achieves an AUROC improvement of .000065 and .060883 over previous models in known and zero-day attack detection, respectively. Finally, the proposed method is extended to open-set recognition achieving OpenAUC improvements of .170883 over existing approaches.\footnote{The implementation and experiments are open-sourced and available at: \url{https://github.com/jackwilkie/CLOSR}}
\end{abstract}


\section{Introduction}
\label{sec:introduction}
\IEEEPARstart{N}{etwork} intrusion detection systems~(NIDS) are a crucial component in any cyber defence strategy. They monitor communications within a network and flag potentially malicious traffic (cyber-attacks) to a CyberSecurity Operations Centre~(CSOC) for further investigation. Machine Learning~(ML) has become the dominant paradigm in NIDS~\cite{Hindy_2020}. When trained on a sufficiently large compilation of examples, ML-based approaches are able to learn the distributions of benign and malicious traffic and accurately discriminate between them~\cite{lycos2017}. However, this approach has its own limitations: they require high volumes of training data, moreover traditional supervised classifiers are able only to identify traffic classes on which they have been trained, and so are vulnerable to zero-day attacks~\cite{Khraisat2019}. 

Methods based on anomaly detection, such as Autoencoders and Support Vector Machines~(SVMs)~\cite{electronics9101684}~\cite{svm2002}, help alleviate this shortcoming of supervised classifiers. These methods operate by learning the distribution of exclusively benign traffic, and flag traffic which significantly differs from the learned distribution as potential intrusions. As these models are trained only on benign data, which is abundantly available, they relieve the burden of collating a significant volume of labelled examples. Furthermore, by training exclusively on benign examples, these models are theoretically equally able to detect well-known and zero-day attacks alike. However, they are also unable to differentiate between new benign traffic subclasses, which leaves them prone high false positive rates, resulting in them being unsuitable for practical applications. While Open-Set Recognition~(OSR) models such as Deep Open Classification~\cite{shu-etal-2017-doc} and OpenMax~\cite{bendale2015opensetdeepnetworks} aim to extend supervised classifiers to also detect novel classes, they have substandard performance in detecting zero-day attacks when applied to NIDS.

Recently, contrastive learning has emerged as a promising approach for NIDS. Methods such as Siamese Networks~\cite{hanan_developing_10.115/3437984.3458842,hanan_leveraging_DBLP:journals/corr/abs-2006-15343} and Triplet Networks have shown promising results, motivated by their performance on imbalanced training data. Network traffic and hence training data is inherently imbalanced: perhaps a fraction of 1\% of which represents an attack. This class imbalance can significantly affect the performance of classically trained machine learning models; contrastive approaches have shown great resilience to imbalanced training data~\cite{hanan_developing_10.115/3437984.3458842}. Despite their ability to train well on highly imbalanced data, like traditional models, contrastive models suffer performance degradation when faced with zero-day attacks~\cite{hanan_thesis}.

In this work, Contrastive Learning for Anomaly Detection~(CLAD) is proposed as a framework capable of modelling the distribution of benign network traffic in embedded space as a von-Mises Fisher (vMF) distribution. Unlike anomaly detectors it learns this distribution by training on both benign and malicious traffic giving it enhanced attack detection at low false-positive rates. As CLAD does not explicitly model the distribution of malicious traffic (as is typical of binary classifiers), it is able to generalise to zero-day attacks. The CLAD framework is then extended to OSR giving the Contrastive Learning for Open-Set Recognition (CLOSR) framework. This is achieved by independently modelling the distribution of each known class in a distinct embedded subspace allowing it to perform multiclass classification. Furthermore, zero-day attacks manifest as vectors orthogonal to the class centroids allowing to them to be easily identified.

The contributions of this work are threefold: (i) a contrastive loss for binary classification that exclusively learns the benign distribution while still training on malicious traffic; (ii) an extension of this loss to open-set recognition, enabling multiclass classification with unknown-class rejection; and (iii) a comprehensive empirical comparison against anomaly detection, supervised, and OSR baselines on Lycos2017, demonstrating consistent gains on both known and zero-day attacks.

\section{Related Works}
\label{sec:related_works}
\subsection{Anomaly Detection}
\label{sec:lit_anomaly}
Anomaly detection involves understanding the typical patterns of a dataset during training and then identifying deviations from this pattern during inference. In NIDS, this approach is used to train a model on the normal behaviour of network traffic, allowing it to recognise any deviations as malicious. Since these models are trained solely on benign data, they can theoretically identify zero-day as well as known attacks. These models can be used as stand-alone classifiers or alternatively can be combined with traditional rules based or signature-based approaches~\cite{Sarker2020, 10.1145/3444370.3444590}.

Distance-based methods represent one of the simplest approaches to anomaly detection. These methods typically calculate the centroid of benign training data and identify anomalies based on their Minkowski distance from this centroid~\cite{doi:10.1137/1.9781611976236.18}. The greater the distance, the more likely a data point is to be an anomaly. Alternative distance measures, such as Frobenius and Grassmannian distances, have also been employed~\cite{10.1007/978-3-319-70087-8_59}. Instead of comparing a samples distance to a centroid, other works have used a samples local outlier factor and the distance to its \emph{K}th nearest neighbour~\cite{electronics9061022}. Most recently, DeepSVDD has attempted to parameterise these methods by learning a mapping from raw feature space to an embedded space in which benign data is constrained to a hypersphere~\cite{pmlr-v80-ruff18a}. However, these methods have been shown to require a large amount of data to be effective~\cite{9239385}. Furthermore, selecting a distance metric can be challenging and choices often don't generalise to zero-day attacks~\cite{9356399}. 

Another approach extends discriminative models to perform anomaly detection. One class SVMs extend SVMs to anomaly detection by learning a boundary which encompasses benign traffic during training. During inference, samples appearing inside this boundary are deemed to be benign; whilst samples outside the boundary are considered to be malicious~\cite{electronics9101684}~\cite{svm2002}. Additionally, Isolation Forests~(IF), an extension of tree-based classification models, can be used to identify anomalies as samples which require many branches to isolate from the training dataset~\cite{10040395}.

Deep learning methods have been extensively utilised for anomaly detection, with numerous studies employing various techniques~\cite{8819688, 8757199}. One common approach involves the use of autoencoders, which take an input and encode it into a lower-dimensional representation, before reconstructing it back to the original dimensionality. The model is trained using a reconstruction objective under the assumption that malicious samples will have a higher reconstruction error at test time~\cite{10.1007/978-3-319-13563-2_27}. 

Several variations of the standard autoencoder have been proposed to enhance performance. Sparse autoencoders remove the bottleneck layer and apply regularisation to prevent the model from learning the identity function~\cite{9239385}. Deep Unsupervised Anomaly Detection~(DUAD) trains an autoencoder while iteratively clustering the data to remove outliers~\cite{li2021deep}. DAE-LR adds a regularisation term to ensure that benign data remains compact in latent space~\cite{nkashama2024deeplearningnetworkanomaly}. Another variant, AutoSVM, applies an SVM to the bottleneck layer to perform classification~\cite{8463474}. Despite the widespread application of autoencoders in NIDS, challenges remain as certain attacks often exhibit low reconstruction errors, which can prevent them from being detected without a high false positive rate~\cite{10.1007/978-3-319-13563-2_27}.

\subsection{Open-Set Recognition}
\label{sec:lit_osr}

Open-Set Recognition is a subset of ML which aims to extend traditional classifiers, which discriminate between a closed-set of known classes, to also identify samples of class distributions which were not represented in the training data. This is applicable to NIDS where it is beneficial for a classifier to be able to identify zero-day attacks in addition to identifying the type of traffic amongst a set of known classes.

One approach to OSR, is to employ a multistage classification framework, which trains both an anomaly detector and closed set classifier. Samples are then either flagged as out of distribution~(OOD) or assigned their predicted closed-set class label based on thresholds set on both the anomaly detector and the classifier's predictions. In NIDS, an autoencoder and random forest have found to be effective for the anomaly detector and classifier, respectively~\cite{10077796}. Deep Open Classification~(DOC) instead trains a classifier for OSR in an end-to-end manner~\cite{shu-etal-2017-doc}. Each known class in the dataset is assigned its own logit and sigmoid activation in the output layer, with each being trained in a one-vs-rest fashion. After training, a Gaussian distribution is fit onto the sigmoid activations for each class, with a sample's likelihood under these distributions being used to detect zero-day attacks at test time.

Another line of work aims to extend pretrained closed-set classifiers to OSR. OpenMax does this by fitting a Weibull distribution to the distances between the activation vectors from the penultimate layer of a neural network and the mean activation vector for each known class~\cite{bendale2015opensetdeepnetworks}. During inference, an OOD score is calculated using the CDFs of the Weibull distributions. CROSR is an extension of OpenMax where a deep hierarchical reconstruction network~(DHRN) is trained instead of a standard MLP~\cite{crosr}. G-OpenMax, another extension of OpenMax trains a neural network with synthesised data as the unknown class~\cite{ge2017generativeopenmaxmulticlassopen}.

Finally, OSR can be performed using statistical analysis in the feature space. Centroid methods compare the features representation to the mean values using a similarity metric~\cite{fei-liu-2016-breaking}. Open set nearest neighbours considers the ratio of the distance from a sample to its nearest class relative to that of the second-nearest class~\cite{osnn}. Weibull SVMs calibrate the predictions of one class SVMs using Weibull distributions~\cite{6809169}. Extreme value machines derive a density function from extreme values theory and use this to model class belongingness probabilities~\cite{Rudd_2018}.

\subsection{Contrastive Learning}
\label{sec:lit_contrastive}

Contrastive learning is a technique in machine learning that trains models to learn an embedded representation of the input data in which similar samples are close together, whilst being separated from dissimilar samples. The simplest form of this technique is exemplified by Siamese Networks, which utilise a contrastive loss function designed to minimise distances between similar pairs and maximise those between dissimilar pairs~\cite{constrastive_loss1}. In NIDS, a Siamese network was initially trained in a supervised fashion, where tabular samples of the same class were treated as similar and different classes were dissimilar~\cite{hanan_developing_10.115/3437984.3458842}. It was later extended to few-shot classification, where the Siamese network was able to identify novel attacks without retraining, given a small number of reference samples~\cite{hanan_leveraging_DBLP:journals/corr/abs-2006-15343}. Since then, contrastive learning has been shown to be effective when dealing with class imbalance~\cite{conflow} and datasets of limited size~\cite{hanan_thesis}.

Due to its effectiveness, contrastive learning has been gaining popularity in NIDS. RENOIR is a contrastive model which employs two autoencoders, one of which has been trained on benign traffic and one which has been trained on malicious traffic. The distance between a sample and its corresponding reconstruction is minimised, whilst that between the sample and the other reconstruction is maximised~\cite{Andresini2021}. FeCo trains an autoencoder using the InfoNCE loss function in a federated learning regime where the model can learn from multiple clients~\cite{9796926}. In online learning regimes contrastive learning has been used to pseudo-label traffic flows, these are then used to periodically retrain the model~\cite{zhang2024aocidsautonomousonlineframework}. Other approaches have used image representations in place of tabular data~\cite{Wang2021}.

Finally, contrastive self-supervised learning~(SSL) approaches are capable of learning features from unlabelled or benign traffic which can then be used for anomaly detection or fine-tuned for classification. SSCL-IDS minimises the distance between two masked views of an input samples whilst maximising it between other samples~\cite{10619725}. Conflow minimises the distance between two views of a sample generated by passing it through a model twice, each time using a different dropout mask~\cite{conflow}. CLDNN trains a contrastive model using feature masking to similar pairs and using other samples as dissimilar pairs~\cite{9935282}.

\section{Proposed Approach}
\label{sec:proposed_method}

\subsection{System Overview}
\label{subsec:approach_overview}

\begin{figure*}[!t]
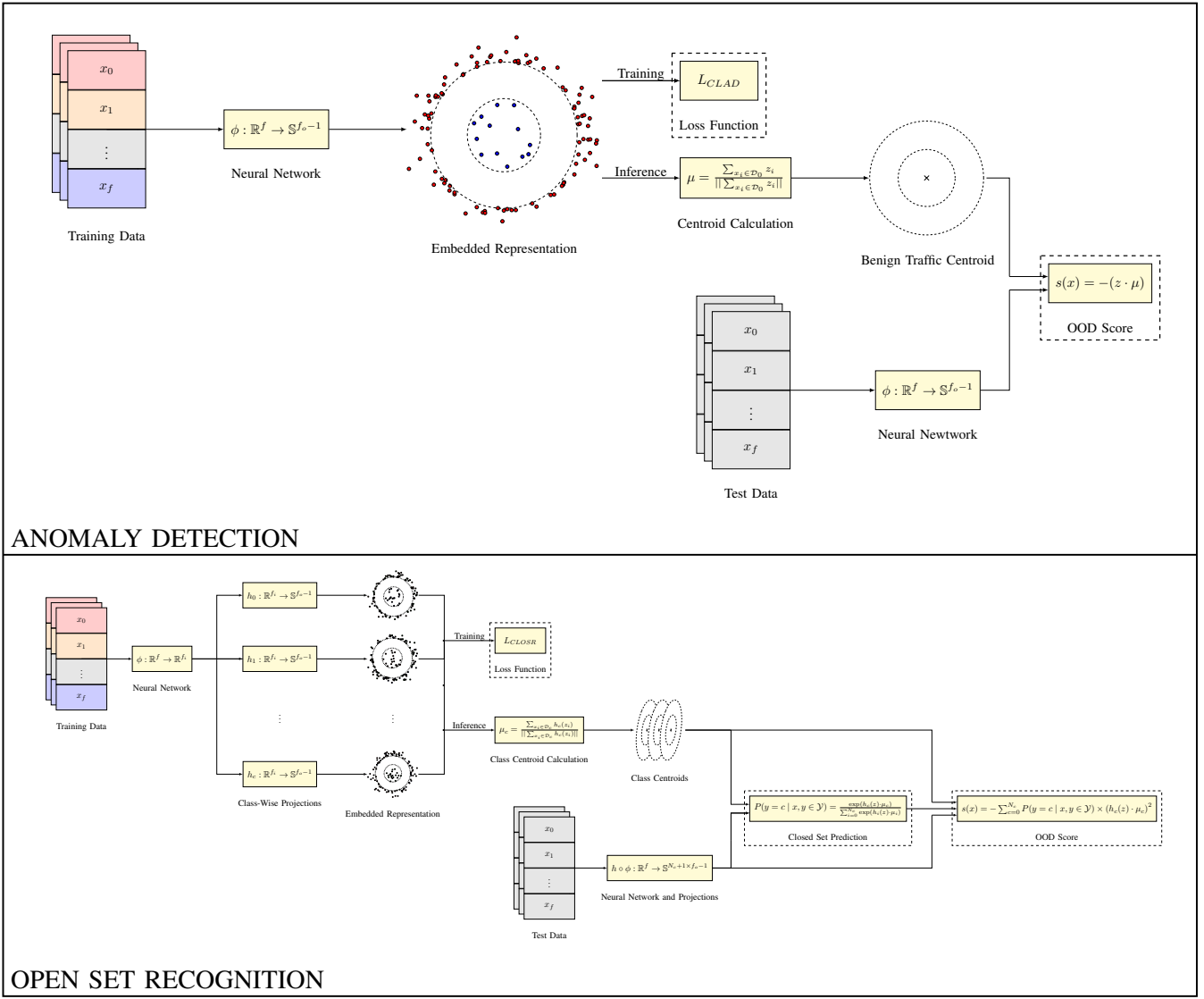

    \centering
    \resizebox{\linewidth}{!}{
        \begin{tikzpicture}

    \node[anchor=south, inner sep=0pt] (clad) {
        \resizebox{\linewidth}{!}{\input{plots/system_overview/anomaly_detection_expanded}}
    };

    \node[anchor=north, inner sep=0pt, yshift=-0.5cm] (closr) at ([yshift=0pt]clad.south) {
        \resizebox{\linewidth}{!}{\input{plots/system_overview/osr_expanded}}
    };

    \draw[thick] 
        ($(clad.north west) + (-0.5, 0.5)$) rectangle 
        ($(closr.south east) + (0.5, -0.5)$);

    \draw[thick] 
        ($(clad.south west) + (-0.5, -0.25)$) -- 
        ($(clad.south east) + (0.5, -0.25)$);

    \node[anchor=south west, font=\large] at ($(clad.south west) + (-0.5, -0.25)$) {ANOMALY DETECTION};

    \node[anchor=south west, font=\large] at ($(closr.south west) + (-0.5, -0.5)$) {OPEN SET RECOGNITION};

\end{tikzpicture}
    }
    \caption{Wholistic overview of the proposed approach. \textbf{Top:} The CLAD loss function learns the distribution of benign traffic as an embedded vMF distribution. During inference the cosine distance between test representations and the centroid of this distribution is used to identify anomalous traffic. \textbf{Bottom:} The CLOSR framework extends CLAD to OSR by modelling the distribution of each known traffic class as an independent vMF distribution in a class specific embedded subspace. Closed set can be inference can then be performed by identifying the distribution with the highest cosine similarity to the test embedding, while zero-day attacks manifest as orthogonal vectors. Here the vMF distributions have been flattened for illustrative purposes.}
    
    \label{fig:system_overview}
\end{figure*}

This section introduces Contrastive Learning for Anomaly Detection (CLAD) and its extension, Contrastive Learning for Open-Set Recognition (CLOSR), designed for detecting zero-day attacks in binary classification and OSR scenarios, respectively. As illustrated in Figure~\ref{fig:system_overview} (top), CLAD trains a neural network to model benign network traffic as a vMF distribution in embedded space. After training, the centroid of this distribution is computed and its cosine distance from test samples is used as a binary classification score at inference time. By leveraging contrastive learning, CLAD avoids explicitly modelling malicious traffic and thus does not make the closed world assumption, despite training on malicious samples. This results in significantly improved known and zero-day attack detection over traditional supervised classifiers and anomaly detectors.

CLOSR, shown in Figure~\ref{fig:system_overview} (bottom), extends CLAD to the OSR setting by introducing class-wise linear projections. These projections are optimised independently using the CLAD loss function, enabling the modelling of each known class as an independent vMF distribution in a distinct embedded subspace. Multiclass classification is then performed by identifying the class centroid with the least cosine similarity to the test embedding. Importantly, zero-day attacks manifest as vectors orthogonal to class centroids, making them easily distinguishable from known classes.

\subsection{Contrastive Learning for Anomaly Detection}
\label{sec:proposed_clad}

In NIDS the objective of machine learning based binary classifiers is to leverage a training dataset to learn decision boundaries between benign and malicious samples, such that malicious traffic can be identified during inference. Concretely, given a training dataset \( \mathcal{D}_{\text{train}} := \{ (x_i,y_i) \mid x \in \mathbb{R}^f, y \in \mathcal{Y} \}_{i=1}^{N_{\text{train}}} \) containing \( N_{\text{train}} \in \mathbb{Z}^+\) recorded network flows---where \(x\) is a set of \(f \in \mathbb{Z}^+\) tabular features representing each flow and \(\mathcal{Y} := \{0,\ldots,N_c\} \) is set of class labels representing benign traffic (\(y = 0\)) and \(N_c \in \mathbb{Z}^+\) known malicious classes of traffic (\(y \neq 0\))--a decision function must be learned to predict the binary class labels \(y' := \mathbbm{1}[y \neq 0] \in \{0, 1\}\) of unknown flows at test time. Here \(\mathbbm{1}\) represents an indicator function which returns 1 when the condition is true and returns 0 otherwise.

The most common solution is to train a parameterised neural network, \(\phi_\theta: \mathbb{R}^f\!\to\!\mathbb{R}\) with parameters \(\theta\), to minimise the binary cross-entropy (BCE) loss function, which is equivalent to maximising the log-likelihood of the conditional distribution \(P(y' \mid x;\theta)\) under a Bernoulli model. In practice, it is calculated over a batch \(\mathcal{B} \subset \mathcal{D}_{\text{train}}\) sampled from the training dataset as shown in Equation~\ref{eq:bce_loss}, where \(\hat{y}_i := \sigma(\phi(x_i))\) is the predicted probability that flow \(x_i\) is malicious and \(\sigma: \mathbb{R} \rightarrow [0,1]\) is the sigmoid activation function. The model parameters \(\theta \) have been omitted for simplicity.

\begin{equation}\label{eq:bce_loss}
     L_{\text{BCE}} = -\frac{1}{|\mathcal{B}|}\sum_{(x_i,y_i) \in \mathcal{B}} \left[ y'_i \log \hat{y}_i + (1 - y'_i) \log (1 - \hat{y}_i) \right]
\end{equation}

NIDS trained using the BCE loss aim to directly model the conditional label distribution \(P(y' \mid x)\). However, when faced with imbalanced data, these models often exhibit poor performance due to implicit biases toward the majority class, reflecting the marginal label distribution \(P(y')\) in practice. Moreover, this approach typically assumes a symmetry between known and zero-day attacks where \(P(y' \mid x,y \in \mathcal{Y}) = P(y' \mid x) \). This symmetry does not hold in practice, limiting the model’s ability to generalise to zero-day attacks.

To circumvent the implicit learning of \(P(y')\), supervised contrastive learning utilises the pairwise interactions between samples to estimate the density ratio \(\frac{P(x \mid y)}{P(x)}\) from which a value proportional to \(P(y \mid x)\) can be recovered under a uniform prior for \(P(y)\). The supervised contrastive loss function is given in Equation~\ref{eq:supcon_loss}~\cite{DBLP:journals/corr/abs-2004-11362}. Here \(\phi: \mathbb{R}^f\!\to\!\mathbb{S}^{f_o - 1}\) is a neural network mapping the data from feature space to an embedded representation  of dimensionality \(f_o \in \mathbb{Z}^+\) such that \(\|\phi(x)\|_2 = 1 \quad \forall x \in \mathbb{R}^f\). \(\mathcal{P}(i) := \{ x_j \in \mathcal{B} \mid j \neq i, y_j = y_i \}\) denotes the set of positive samples for anchor \(x_i\), and \(\mathcal{A}(i) := \{ x_j \in \mathcal{B} \mid j \neq i \}\) denotes all other samples in the batch. Finally, \( \tau \in \mathbb{R}^+ \) is the temperature hyperparameter. The notation \( z_i := \phi(x_i) \) is used for simplicity.

\begin{align} \label{eq:supcon_loss}
     L_{\text{SupCon}} &= \frac{1}{|\mathcal{B}|} \sum_{(x_i,y_i) \in \mathcal{B}} \frac{1}{|\mathcal{P}(i)|} 
     \sum_{x_p \in \mathcal{P}(i)} 
     \notag\\&-\log \left( 
          \frac{\exp\left(z_i \cdot z_p / \tau\right)}{\displaystyle \sum_{x_a \in \mathcal{A}(i)} \exp\left(z_i \cdot z_a / \tau\right)
          }
     \right)     
\end{align}

While contrastive learning has achieved improved performance in NIDS tasks due to its ability to model imbalanced data; existing methods maintain the symmetry assumption between known and unknown malicious classes and thus still suffer from reduced performance when faced with zero-day attacks. This work instead proposes the CLAD framework, which learns the density ratio \(\frac{P(x \mid  y=0)}{P(x \mid y \neq 0)}\) while explicitly modelling \(P(x \mid y=0)\) as a vMF distribution, which has been shown in computer vision literature to be well suited for OOD detection~\cite{ming2023exploithypersphericalembeddingsoutofdistribution}. As the density ratio is asymmetrically computed over benign traffic, the model exclusively learns the distribution of benign traffic, with malicious samples being used only to normalise the benign distribution. This relaxes the closed world assumption akin to anomaly detectors (i.e. \( P(x \mid y' = 0, y\in \mathcal{Y}) =  P(x \mid y' = 0)\)). Furthermore, by explicitly modelling the data as vMF distributions, a structure is forced on embedded representation of the data allowing for zero-day attacks to be detected.

Additionally, it should be noted that when a shared embedding space is used for many classes, the optimal arrangement of class centroids on the hypersphere approaches a regular simplex configuration, in which the pairwise cosine similarity is fixed at \(\frac{1}{N_c + 1}\). This optimal packing reduces the angular separation between centroids as the number of classes increases, which in turn lowers the margin between known and zero-day samples. By contrast, learning the distribution of only benign traffic maximises the margin between benign samples and zero-day attacks. 

The CLAD loss function is derived by minimising the negative log-likelihood of benign samples under the density ratio \( \frac{P(x \mid y=0)}{P(x \mid y\neq 0)} \), using vMF class-conditional distributions as shown in the loss function given in Equation~\ref{eq:density}. Here \(\mu_0 \in \mathbb{S}^{f_o -1} \) and \(\mu_{\neg 0} \in \mathbb{S}^{f_o -1} \) are the centroids of the class-conditional distributions \(P(x \mid y=0)\) and \( P(x \mid y \neq 0) \), respectively. The vMF normalisation terms \(C_{f_o}(\kappa_0)\) and \(C_{f_o}(\kappa_1)\) have been omitted due to forming the additive constant \(-\log(\frac{C_{f_o}(\kappa_0)}{C_{f_o}(\kappa_{1})})\), which is eliminated upon differentiation. The notations \(\mathcal{B}_c := \{ x_j \in \mathcal{B} \mid y_j = c \}\) and \(\mathcal{D}_c := \{ x_j \in \mathcal{D}_{\text{train}} \mid y_j = c \}\) are used to represent the subset of samples in the batch and training dataset belonging to class \(c \in \mathcal{Y}\), respectively. Finally, \(d: \mathbb{S}^{f_o - 1} \times \mathbb{S}^{f_o - 1}\!\to\![0,1]\) is the rescaled cosine distance function shown in Equation~\ref{eq:cosine_dist} for two unit vectors \(z\) and \(z'\).

\begin{equation} \label{eq:density}
     L = \frac{-1}{|\mathcal{B}_0|}
     \sum_{x_i \in \mathcal{B}_0} 
     \log\left(
         \frac{
             \exp\left(-\kappa_0 d(z_i, \mu_0)\right)
         }{
          \exp\left(-\kappa_1 d(z_i, \mu_{\neg 0} )\right)
         }
     \right)
     \end{equation}
     
\begin{equation}\label{eq:cosine_dist}  
     d(z,z') = \frac{1-(z \cdot z')}{2}
\end{equation}

To circumvent the need for computing centroids over the entire training dataset during optimisation, the relationship between an embedding and samples drawn from a vMF distribution, as defined in Equation~\ref{eq:vmf_expectation}, is leveraged. This relationship enables the centroids of the vMF distributions to be approximated via Monte Carlo sampling using samples within the current batch, as illustrated in Equation~\ref{eq:density_monte_carlo}, where \(\mathcal{N}(i) := \{ x_j \in \mathcal{B} \mid j \neq i, y_j \neq y_i \}\) denotes the set of negative pairs for anchor \(x_i\) and \( \kappa_{y'} \in [0,\inf)\) is class-specific concentration parameter of vMF distribution associate with binary class label \(y'\).

\begin{equation}\label{eq:vmf_expectation}
     d(z, \mu) \propto \mathbb{E}_{z' \sim \mathrm{vMF}(\mu, \kappa)}\left[d(z, z')\right]
\end{equation}     

\begin{equation} \label{eq:density_monte_carlo}
     L = \frac{-1}{|\mathcal{B}_0|} 
     \sum_{x_i \in \mathcal{B}_0} 
     \log\left(
         \frac{
             \exp\left(\frac{-\kappa_0}{|\mathcal{P}(i)|}\sum\limits_{x_p \in \mathcal{P}(i)} d(z_i, z_p)\right)
         }{
             \exp \left(\frac{-\kappa_1}{|\mathcal{N}(i)|}\sum\limits_{x_n \in \mathcal{N}(i)} d(z_i, z_n)\right)}
     \right)
\end{equation}

Finally, simplifying Equation~\ref{eq:density_monte_carlo} gives the final CLAD loss function given in Equation~\ref{eq:clad_loss}. Here the concentration terms have been tied such that \(\kappa_0 = \kappa_1\) and removed as a multiplicative constant, which is absorbed into the learning rate during hyperparameter optimisation. In this formulation, the distance terms are squared to enhance the separation between benign and malicious traffic in the embedded space. 

\begin{align}
     L_{CLAD} &= \frac{1}{|\mathcal{B}_0|}
         \sum_{x_i\in\mathcal{B}_0} \Bigg[
             \frac{1}{|\mathcal{P}(i)|} 
             \sum_{x_p \in \mathcal{P}(i)} d(z_i, z_p)^2 \nonumber \\
     & + \sum_{x_n \in \mathcal{N}(i)} 
     \frac{1}{|\mathcal{N}(i)|}\left(1 - d(z_i, z_n)\right)^2
         \Bigg]
\label{eq:clad_loss}
\end{align}

The CLAD loss is reminiscent of the hinge-based contrastive loss function~\cite{constrastive_loss1}, however, differs in several fundamental ways. First, CLAD is evaluated exclusively on benign anchor samples, whereas the contrastive loss optimises learns the distribution of each known class symmetrically. This removes the closed-world assumption inherent in contrastive learning and aligns CLAD with anomaly-style training objectives. Second, CLAD can be viewed as fixing the contrastive loss' margin term to \(m=1.0\) which forces benign and malicious traffic to converge to antipodal regions of the hypersphere and eliminates the hinge regularisation. Third, unlike the geometric motivation of contrastive loss, CLAD admits a probabilistic interpretation via a vMF likelihood ratio, yielding a density-ratio estimator rather than a geometric separation loss

It is pertinent to note that squaring the distance metric preserves the bound range of the rescaled distance metric i.e. \(d^2: \mathbb{S}^{f_o - 1}\times \mathbb{S}^{f_o - 1}\!\to\![0,1]\). By squaring the distance metric the gradient magnitudes are scaled by factors of \(2|d(z,z')|\) and \(2|d(z,z') - 1|\) for positive and negative pairs, respectively. This results in larger gradients for positive pairs with large separations and for negative pairs with poor separation, giving a more discriminative embedded representation. This is similar to the temperature scaling used by the SupCon loss function~\cite{DBLP:journals/corr/abs-2004-11362}, where sharper gradients improve class separability. 

\subsection{Outlier Detection}
\label{sec:proposed_inference}

CLAD models benign network traffic as a vMF distribution in embedded space. This allows for inference to be performed by evaluating the likelihood of a test embedding under the learned benign vMF distribution. To facilitate this the mean direction, \( \mu_0 \in \mathbb{S}^{f_o - 1} \), of the benign distribution is calculated as the centroid of benign data, normalised to lie on the unit hypersphere as shown in Equation~\ref{eq:centroid}.

\begin{equation}\label{eq:centroid}
     \mu = \frac{\sum_{x_i \in \mathcal{D}_0} z_i}{||\sum_{x_i \in \mathcal{D}_0} z_i||}
\end{equation}

The cosine similarity between a test embedding and the distribution centroid is proportional to the likelihood \( P(x \mid y=0)\). In principle, this can be converted to a conditional probability \(P(y=0 \mid x)\) using Bayes' theorem; however, computing the marginal likelihood \(P(x)\) requires integrating over all possible class-conditional distributions, which is only possible under a closed world assumption. Thus, CLAD instead uses the negative cosine similarity between a test embedding and the centroid as an OOD score, \(s(x) \in [-1,1]\), given by Equation~\ref{eq:benign_likliehood}. This score is proportional to the negative log-likelihood of the sample under the vMF distribution, up to an additive constant. This is in line with prior work in OOD detection for computer vision~\cite{DBLP:journals/corr/abs-2010-03759}, where log-likelihoods are widely used as effective OOD scores.
 
\begin{equation}\label{eq:benign_likliehood}
     s(x) = - (z \cdot \mu )
\end{equation}

Binary classification can be then made by placing a threshold, \( \tau \in [-1,1]\) on the OOD score. As shown in Equation~\ref{eq:anomaly_pred_label}: test samples with an OOD score less than this threshold are deemed to malicious, while those with an OOD score less than or equal to the threshold are considered benign. It is important to note that this work employs threshold independent metrics such as AUROC and FPR@95 and thus does not calibrate the value of the threshold. In practical scenarios the network's CSOC would determine an acceptable false positive rate based on available resources and select a corresponding value using a held-out validation set of benign traffic.

\begin{equation} \label{eq:anomaly_pred_label}
     \hat{y} = 
     \begin{cases}  
          0, & \text{if } s(x) < \tau; \\
          1, & \text{otherwise}.
     \end{cases}
\end{equation}

\subsection{Extension to Open-Set Recognition}
\label{sec:proposed_multiclass}

The CLAD loss function learns to discriminate the distribution of benign traffic from other classes, resulting in binary predictions indicating whether the unknown sample is benign or malicious. In practical scenarios, it is often beneficial not only to know that traffic is malicious, but also the type of attack being performed. To achieve this, CLAD is first modified to learn the distribution of an arbitrary class, \(c \in \mathcal{Y}\), as shown in Equation~\ref{eq:clad_linear_heads}. To facilitate this, the neural network, \( \phi: \mathbb{R}^f\!\to\!\mathbb{R}^{f_i} \), projects the data to an intermediate representation, \(z_i = \phi(x_i)\), which is then mapped to distinct unit hyperspheres using class-wise linear projection heads, \( \{h_c: \mathbb{R}^{f_i}\!\to\!\mathbb{S}^{f_o -1} \}_{c=0}^{N_c} \).  These heads are independently optimised to map samples onto a unique hypersphere, where its associated class distribution is modelled using a vMF distribution.

\begin{align}
     &L_{CLAD}(c) = \frac{1}{|\mathcal{B}_c|}
         \sum_{x_i\in\mathcal{B}_c} \Bigg[
             \frac{1}{|\mathcal{P}(i)|} 
             \sum_{x_p \in \mathcal{P}(i)} d(h_c(z_i), h_c(z_p))^2 \nonumber \\
     & + \sum_{x_n \in \mathcal{N}(i)} 
     \frac{1}{|\mathcal{N}(i)|}\left(1 - d(h_c(z_i), h_c(z_n))\right)^2
         \Bigg]
\label{eq:clad_linear_heads}
\end{align}

The CLOSR loss function can then be defined as the CLAD loss computed over each class, where each class employs its distinct linear projection. The CLOSR loss is formally described by Equation~\ref{eq:closr_loss}. 

\begin{equation}
     L_{CLOSR} = \sum_{c \in \mathcal{Y}} L_{CLAD}(c)\label{eq:closr_loss}
\end{equation}

After training, a mean direction (centroid) is computed for each class, $\mu \in \mathbb{S}^{N_c + 1 \times f_o -1}$; where $\mu_c$ is the centroid of class \emph{c}. As shown in Equation~\ref{eq:closr_centroids}, this is calculated as the mean of all embeddings of a given class in the training data, in its respective linear projection, and then normalised to lie on a unit hypersphere.

\begin{equation}\label{eq:closr_centroids}
     \mu_{c} = \frac{\sum_{x_i \in \mathcal{D}_c} h_c(z_i)}{||\sum_{x_i \in \mathcal{D}_c} h_c(z_i)||}
\end{equation}

By comparing the cosine similarity between an unknown test embedding and the centroid of each class's vMF distribution, unnormalised class-conditional likelihoods \( P(x \mid y=c)\) can be computed for each class. These can be converted to a closed-set prediction under a uniform prior with Bayes' theorem using the softmax function as shown in Equation~\ref{eq:closr_softmax}.

\begin{equation}\label{eq:closr_softmax}
     P(y = c \mid x, y\in \mathcal{Y}) = \frac{\exp\left(h_c(z)\cdot \mu_c\right)}
     {\sum_{i \in \mathcal{Y}} \exp\left(h_i(z) \cdot \mu_i\right)}
\end{equation}

As the model is not explicitly trained on zero-day attacks, these samples can be assumed to be more uniformly distributed over the unit hypersphere than those from known attack classes. Under this assumption, zero-day embeddings exhibit many weakly correlated dimensions, allowing the cosine similarity between a zero-day sample and a class mean to be approximated as a sum of independent random variables, each weighted by the corresponding component of the class centroid. By the Central Limit Theorem, this sum converges to a Gaussian distribution with zero mean and variance \(1/f_o\), as formalised in Equation~\ref{eq:ctl_expectation}.

\begin{equation} \label{eq:ctl_expectation}
     h_c(z) \cdot \mu_c = \sum_{i=1}^{f_o} h_c(z)_i \mu_{c,i} \sim \mathcal{N}\left(0, \frac{1}{f_o} \right) 
\end{equation}

Following Equation~\ref{eq:ctl_expectation}, an OOD score, \(s(x) \in [-1,0]\), is defined in Equation~\ref{eq:ood_score}. This is proportional to the log-likelihood of a test embedding belonging to the Gaussian distribution up to an additive constant. A high score is expected to indicate a zero-day attack. 

\begin{equation} \label{eq:ood_score}
     s(x) = - \sum^{N_c}_{c=0} P(y = c \mid x, y \in \mathcal{Y}) \times \left( h_c(z) \cdot \mu_c \right)^2
\end{equation}

While this weighted sum assumes isotropy and independence amongst the cosine similarity measurements; these assumptions are explicitly supported in this setting due the construction of the embedding space and the properties of the hyperspherical distributions. Isotropy is ensured under the assumption of zero-day embeddings being uniformly distributed over the unit hypersphere. In this case, the projection of a random unit vector onto any fixed unit direction follows a Gaussian distribution with zero mean and constant variance \(1/f_o\), regardless of the class. Furthermore, the independence assumption is supported by the architecture: each class-wise linear projection head is independently optimised using a one-vs-rest strategy allowing for them to be treated as independent. This is a common assumption made by ensemble modelling and mixture of experts architectures~\cite{10.1007/3-540-45014-9_1},~\cite{Rifkin2004InDO},~\cite{puigcerver2024sparsesoftmixturesexperts}.

Finally, a threshold, $\tau \in [-1,0]$, is placed on the OOD score. As shown in Equation~\ref{eq:closr_label_prediction}, samples with an OOD score above the threshold are predicted to be the closed-set class with the highest predicted probability; whilst those with an OOD score equal to or less than the threshold are predicted as belonging to a previously unseen class ($\hat{y} = -1$).

\begin{equation}
    \hat{y} = 
    \begin{cases} 
        -1, & \text{if } s(x) > \tau; \\
        argmax(P(y \mid x, y\in \mathcal{Y})), & \text{otherwise}.
    \end{cases}
    \label{eq:closr_label_prediction}
\end{equation}

It should be noted that, similar to the binary classification threshold, the threshold \(\tau\) is not calibrated in this work due to the use of threshold independent evaluation metrics. In practical deployments, this would be tuned on a held-out closed-set validation set based on acceptable error rates defined by the corresponding cybersecurity operations centre~(CSOC).

The CLAD loss function is a special case of the CLOSR loss. Given an \(N = N_c + 1\) way classification task, CLOSR explicitly models \(N_c\) known class distributions with the remaining class being treated as OOD samples. For an \(N = 2\) way classification task, a single benign class distribution is learned (with known malicious and OOD traffic being considered a single class), yielding the CLAD training objective exactly. 

\subsection{Network Architecture}
\label{sec:proposed_network}

The CLAD and CLOSR loss functions are parameterised by a Multi-Layer Perceptron~(MLP), notated as \(\phi\). The proposed architecture initially projects the input feature vector to a higher-dimensional space of dimensionality \(d_{model} \in \mathbb{Z^+}\) using a linear transformation. Subsequently, a non-linear function is learned through a series of sequential MLP blocks. Each block comprises a linear transformation, which maintains the width \(d_{model}\), followed by a ReLU activation function. By using a fixed width across all MLP blocks, the model architecture is simplified, requiring only a single width instead of a distinct width for each layer.

Finally, a linear projection head is used to downsample the data to its embedded dimensionality and project it onto a unit hypersphere. While the CLAD loss function employs a single projection head, the CLOSR loss function introduces multiple projection heads in order to facilitate the training of class-wise subspaces in a one-vs-rest manner. It should be noted that while the CLAD loss function introduces sampling noise through its Monte Carlo approximation, training remained stable without introducing normalisation or variance reduction techniques to the architecture.

\section{Comparison to Baseline Models}
\label{sec:exp_results}

\subsection{Experimental Procedure}
\label{sec:exp_procedure}

This work trains and evaluates models using the Lycos2017 dataset~\cite{lycos2017}, which contains 1,789,954 flows across 14 classes of traffic, including both malicious and benign samples. The dataset is highly imbalanced, with benign traffic consisting of over 1,000,000 samples, and malicious classes ranging from as few as 11 samples to as many as 100,000 samples. The dataset was split into a 50\%-50\% train-test sets, ensuring that each set contained an even number of samples of each class. The Web Attack~(SQL Injections) and Heartbleed classes were an exception, as they were only contained in test set to simulate zero-day attacks.

In this section CLAD and CLOSR are evaluated and compared to baseline models from anomaly detection, supervised classification, and OSR. To ensure fair evaluation, the proposed approaches and baseline models were optimised and evaluated using an identical procedure: machine learning models were trained for 200 epochs using the AdamW optimiser~\cite{loshchilov2019decoupledweightdecayregularization}. The learning rate was scheduled using a linear warmup for 20 epochs followed by cosine annealing~\cite{loshchilov2017sgdrstochasticgradientdescent}. Weighted class balancing was also used to sample batches. The base learning rate, batch size, weight decay, dropout rate and the model's width and depth were tuned as hyperparameters. The cosine distance metric was used for all models.

The hyperparameters of gradient based approaches were optimised using 200 iterations of random search, with each iteration employing 5-fold of cross-validation across the training data. Here  200 iterations were used for each model due to the available compute budget. Standard search ranges were used for each parameter, with search ranges being kept constant across models. Model specific hyperparameters were searched across parameter ranges suggested in the original works. For gradient-free approaches it was feasible to perform hyperparameter optimisation using a grid search across typical hyperparameter ranges. For fair comparison with the contrastive loss function, CLAD and CLOSR were trained using an equivalent margin and hinge regularisation term allowing for an identical search space, however, all optimised models converged to a margin value of \(m=1.0\).

After hyperparameter optimisation, the configuration with the greatest mean AUROC across each cross-validation fold was selected for training on the entire training dataset and evaluation on the test set. Mean test set results over 20 training and evaluation runs are reported. Additionally, statistical testing was performed between CLAD/CLOSR and the leading baseline model using a Wilcoxon signed-rank test, with statistical significance being indicated where the p-value was below the 0.01 threshold.

\subsection{Comparison to Binary Classification Models}
\label{sec:exp_anomaly_eval}

\begin{table*}[!t] 
    \caption{AUROC comparison of CLAD and baseline models when detecting known (top) and zero-day (bottom) attacks. CLAD significantly outperforms supervised classifiers and anomaly detectors in zero-day attack detection. Best performance is shown in \textbf{bold} and * indicates statistical significance.} 
    \label{tab:binary_results_auroc} 
    \centering 
    \resizebox{\textwidth}{!}{
    \begin{tabular}{cccccccccccc} 
    \toprule
    
    Class & CLAD & DUAD & DAE-LR & Deep SVDD & AE & SVM & AutoSVM & IF & RENOIR & MLP & Siamese Network \\

    
    \midrule
    Botnet & .999980 & .888620 & .720118 & .626321 & .648185 & .637679 & .662214 & .696588 & .999990 & .999982 & .999992 \\

    DDoS & .999995 & .987811 & .998323 & .971817 & .957563 & .889780 & .722085 & .778166 & .999996 & .999995 & .999881\\
    
    DoS (Golden Eye) & .999585 & .930244 & .954869 & .740828 & .878592 & .846840 & .796121 & .727087 & .999425 & .999397 & .998539 \\

    DoS (Hulk) & .999995 & .988089 & .993563 & .899340 & .950815 & .894898 & .793412 & .781731 & .999991 & .999985 & .999778 \\

    DoS (Slow HTTP Test) & .999575 & .911725 & .975763 & .659927 & .978112 & .963021 & .925308 & .848820 & .999831 & .999543 & .999474 \\

    DoS (Slow Loris) & .999876 & .947156 & .980669 & .692552 & .938116 & .896824 & .776275 & .775622 & .999928 & .999937 & .999826 \\

    FTP Patator & .999986 & .982829 & .952138 & .822762 & .779748 & .736828 & .702073 & .759742 & .999991 & .999987 & .999948 \\

    Portscan & .999993 & .943170 & .974972 & .732919 & .871651 & .741269 & .584936 & .742066 & .999986 & .999962 & .999823 \\

    SSH Patator & .999930 & .961050 & .961583 & .671834 & .824424 & .799072 & .789865 & .816086
    & .999951 & .999980 & .999978 \\

    Web Attack (Brute Force) & .999754 & .836997 & .731959 & .635410 & .801017 & .767771 & .744257 & .748669 & .999312 & .998011 & .996850 \\

    Web Attack (XSS) & .999732 & .909319 & .773056 & .640942 & .801203 & .761677 & .734927 & .768263 & .999286 & .998353 & .996466 \\
    
    \textbf{Closed Set Mean} & \textbf{.999855*} & .935183 & .910637 & .735877 & .857221 & .812333 & .748316 & .767531 & .999790 & .999557 & .999141 \\
    
    \midrule

    Heartbleed  & .995557 & .987390 & .999798 & .985815 & .995161 & .993468 & .988778 & .955030 & .780736 & .071957 & .692232 \\
    
    Web Attack (SQL Injection) & .997696 & .884098 & .777752 & .717808 & .686157 & .745087 & .767333 & .721814 & .991721 & .995762 & .997427 \\

    \textbf{Open Set Mean} & \textbf{.996627*} & .935744 & .888775 & .851812 & .840659 & .869277 & .767333 & .838422 & .886228 & .533859 & .844829 \\

    \bottomrule
    \end{tabular}
    }
    \end{table*}
    
\begin{table*}[!t] 
    \caption{FPR@95 comparison of CLAD and baseline models when detecting known (top) and zero-day (bottom) attacks. CLAD significantly outperforms supervised classifiers and anomaly detectors in zero-day attack detection. Best performance is shown in \textbf{bold} and * indicates statistical significance.} 
    \label{tab:binary_results_fpr} 
    \centering 
    \resizebox{\textwidth}{!}{
    \begin{tabular}{cccccccccccc} 
    \toprule

    Class & CLAD & DUAD & DAE-LR & Deep SVDD & AE & SVM & AutoSVM & IF & RENOIR & MLP & Siamese Network \\
    \midrule


    Botnet & .000020 & .133028 & .334342 & .705342 & .382650 & .393574 & .367771 & .732462 & .000015 & .000024 & .000011 \\

    DDoS & .000009 & .023125 & .004296 & .155205 & .068373 & .170113 & .323883 & .586338 & .000009 & .000008 & .000157 \\

    DoS (Golden Eye) & .001110 & .187053 & .151281 & .750513 & .309403 & .252365 & .247377 & .554949 & .001085 & .002366 & .001829 \\

    DoS (Hulk) & .000011 & .025627 & .014094 & .192490 & .068418 & .144633 & .230494 & .433950 & .000020 & .000051 & .000283 \\
     
    DoS (Slow HTTP Test) & .000818 & .181274 & .038792 & .867094 & .081772 & .181849 & .231434 & .441208 & .000335 & .000738 & .000819 \\

    DoS (Slow Loris) & .000036 & .202757 & .074781 & .860426 & .244468 & .365951 & .604420 & .702358 & .000091 & .000036 & .000175 \\

    FTP Patator & .000013 & .017932 & .051922 & .493948 & .221201 & .263549 & .299008 & .664956 & .000006 & .000004 & .000057 \\

    Portscan & .000008 & .125238 & .051744 & .702921 & .175783 & .306717 & .474476 & .634320
    & .000013 & .000016 & .000218 \\

    SSH Patator (Brute Force) & .000112 & .042739 & .042120 & .722980 & .185131 & .204481 & .211120 & .420270 & .000057 & .000031 & .000029 \\

    Web Attack (Brute Force) & .000295 & .923579 & .795793 & .699538 & .315095 & .264761 & .275640 & .671678 & .000813 & .003889 & .005108 \\

    Web Attack (XSS) & .000298 & .923851 & .795767 & .682162 & .206077 & .246975 & .273129 & .600777 & .000831 & .003952 & .005132 \\
    
    \textbf{Closed Set Mean} & \textbf{.000248} & .253291 & .214085 & .621147 & .205306 & .254088 & .321705 & .585752 & .000298 & .001010 & .001256 \\
    
    \midrule

    Heartbleed  & .007264 & .019323 & .000848 & .094894 & .022762 & .067708 & .105113 & .110316 & .339468 & .993871 & .326209 \\

    Web Attack (SQL Injection) & .003331 & .218930 & .343934 & .601108 & .346540 & .286927 & .256526 & .657605 & .027203 & .005510 & .006520 \\
    
    \textbf{Open Set Mean} & \textbf{.005297*} & .119126 & .172391 & .348001 & .184651 & .177318 & .180819 & .383960 & .183336 & .499691 & .166364 \\

    \bottomrule
    \end{tabular}
    }
    \end{table*}

The efficacy of CLAD when compared to baseline models in binary classification was primarily evaluated by measuring the model's area under the ROC curve (AUROC). This metric represents the expectation of the model assigning a higher classification score to a malicious sample than to to a benign one, thereby quantifying the trade-off between true positive and false positive rates. The results are given in Table~\ref{tab:binary_results_auroc} (top) for known attack classes and Table~\ref{tab:binary_results_auroc} (bottom) for zero-day attack classes. Similarly, the FPR@95 metric, which is the model's false positive rate when classifying an attack class with a 95\% recall, is reported for CLAD and baselines models for known traffic in Table~\ref{tab:binary_results_fpr} (top) and for zero-day attacks in Table~\ref{tab:binary_results_fpr} (bottom). 

On known attack classes, anomaly detectors expectedly demonstrated poor performance when compared to supervised classifiers. As such, CLAD was found to significantly outperform anomaly detectors on known attacks. When compared to supervised classifiers CLAD achieved statistically significant performance improvements in AUROC. While CLAD outperformed supervised classifiers on known attacks in FPR@95; this improvement was not found to be statistically significant.

On zero-day attack detection, supervised classifiers suffer from significant performance degradation, indicating overfitting on known traffic. In contrast, anomaly detectors have circumvented this issue by training only on benign traffic, consequently outperforming the supervised classifiers. Remarkably, despite training on malicious samples, CLAD maintains its strong performance against zero-day attacks. By leveraging the representations learned from both benign and malicious traffic, CLAD surpasses all baselines with statistically significant performance improvements on both AUROC and FPR@95. These results highlight the applicability of CLAD to the binary classification of both known and zero-day attack classes.

\subsection{Comparison to Open-Set Recognition Models}
\label{sec:exp_osr_eval}

\begin{table}[!t] 
    \renewcommand{\arraystretch}{1.3}
    \caption{Comparison of the proposed CLOSR framework with baseline models in OSR. CLOSR outperforms existing models in both open set AUC and OpenAUC. Best performance is shown in \textbf{bold} and * indicates statistical significance.} 
    \label{tab:osr_results} 
    \centering 
    \resizebox{\linewidth}{!}{
    \begin{tabular}{cccc} 
    \toprule
    
    Model & Closed-Set Acc & Open-Set AUC & OpenAUC\\
    
    \hline

    CLOSR & .995276 & \textbf{.974022*} & \textbf{.969420*} \\
    MultiStage & .996612 & .801251 & .798537 \\
    DOC & .995536 & .570263 & .567717 \\
    OPENMAX & .995615 & .720174 & .717016 \\
    CRSOR & .994940 & .748295 & .744509 \\
    Siamese Network & .\textbf{997722*} & .720811 & .719167 \\
    \midrule
    CLOSR + Siamese Network & .997722 & .974022 & .971803 \\
    \bottomrule
    \end{tabular}
    }
    
    \end{table}
\begin{table}[!t] 
    \renewcommand{\arraystretch}{1.3}
    \caption{Closed set classification performance comparison between CLOSR and baseline models. Best performance is shown in \textbf{bold} and * indicates statistical significance.} 
    \label{tab:closed_set_results} 
    \centering 
    \resizebox{\linewidth}{!}{
    \begin{tabular}{cccccc} 
    \toprule
    
    Model & Precision & Recall & PR-AUC & F1 Score & FP Rate\\

    \hline
    CLOSR & .818063 & .929404 & .870178 & .853196 & .005340 \\
    MultiStage & \textbf{.972191*} & .884658 & .140567 & .889541 & .001893 \\
    DOC & .832814 & \textbf{.940472*} & \textbf{.873167*} & .861716 & .004897 \\
    OPENMAX & .835872 & .937882 & .865545 & .861687 & .004746 \\
    CRSOR & .821883 & .938760 & .830984 & .853783 & .005658 \\
    Siamese Network & .910641 & .897597 & .855822 & \textbf{.903406*} & \textbf{.000712*} \\
    \bottomrule
    \end{tabular}
    }
    
    \end{table}

The OSR performance of CLOSR was benchmarked against a range of baseline models, including state-of-the-art approaches from both the NIDS and OSR literature. Models were evaluated using closed-set accuracy, open-set AUC, and OpenAUC~\cite{NEURIPS2022_9f73d65a}, where OpenAUC is defined as the product of the former two metrics. In this context, open-set AUC quantifies the probability of the model correctly distinguishing test samples from unknown classes against those from the known class distributions. The results, reported in Table~\ref{tab:osr_results}, highlight the strong open-set performance of CLOSR. Notably, CLOSR achieves a substantially higher open-set AUC than all baseline methods. While this improvement is accompanied by a decrease in closed-set classification accuracy; CLOSR achieves the highest overall performance by a significant margin, as reflected in its superior OpenAUC score. 

It should be noted that CLOSR can be used as an independent out of distribution detector and paired with a supervised classifier which performs closed-set classification. In this regime, the combination of the CLOSR and the Siamese network achieves the open-set performance of CLOSR and the closed-set performance of the Siamese network, matching the best performance in both open and closed-set classification. For completeness, additional multiclass classification metrics are reported in Table~\ref{tab:closed_set_results}. These results are consistent with the above findings: CLOSR exhibits slightly lower closed-set performance compared to baselines, yet its effectiveness in open-set scenarios makes it the best OSR model in practice.

\section{Additional Results and Ablations}
\label{sec:ablations}

\subsection{Experimental Procedure}
\label{sec:additional_procedure} 

In this Section results from additional experiments and ablation studies are reported. To facilitate this, the dataset was split into training and test sets identical to those used in Section~\ref{sec:exp_results}. To avoid data leakage from the zero-day classes, the training data was further divided 80\%-20\% into training and validation sets, with models being trained on the training data, and validation performance being reported. 

In order to enable fair comparison between CLAD and the contrastive loss function, and to ensure that ablations were unbiased by the design choices used during hyperparameter optimisation, models were trained using a fixed architecture and set of commonly used hyperparameters. An exception was made for Section~\ref{subsec:ablations_copmute} which reports the computational requirements of the optimised models.

\subsection{CLAD Loss Ablations}
\label{sec:additional_hp_sensitivity}

\begin{figure}[!t]
    \centering
    \resizebox{0.7\linewidth}{!}{
    \begin{tikzpicture}
    \begin{axis}[
        xlabel={Margin},
        ylabel={AUROC},
        xmin=0.1, xmax=1,
        ymin=0.99, ymax=1.,
        xtick={0.0, 0.1, 0.2, 0.3, 0.4, 0.5, 0.6, 0.7, 0.8, 0.9, 1.0},
        axis lines=left,
        axis line style={-},
        grid=none,
        ymajorgrids=true,
        legend pos=south east,
        legend style={
            draw=gray!50, 
            line width=.1pt, 
        },
        major grid style={line width=.1pt, draw=gray!25},
        scaled y ticks=false, 
        yticklabel style={/pgf/number format/fixed, /pgf/number format/precision=6}
    ]
    
    \addplot[
        color=orange,
        mark=*,
        line width=1pt
        ]
        coordinates {
            (0.100000,0.991756158879226)(0.200000,0.995471973423135)(0.300000,0.998144289814212)(0.400000,0.997988176604819)(0.500000,0.999518991376386)(0.600000,0.999477773058838)(0.700000,0.999732426764724)(0.800000,0.999757760011855)(0.900000,0.999789944119108)(1.000000,0.999774241963903)
        };
        \addlegendentry{Squared}

    \addplot[
        color=blue,
        mark=*,
        line width=1pt
        ]
        coordinates {
            (0.100000,0.998123976716771)(0.200000,0.995580888173669)(0.300000,0.997530547083871)(0.400000,0.996389573280707)(0.500000,0.999542301057855)(0.600000,0.997593407801701)(0.700000,0.998366621361798)(0.800000,0.996924054818495)(0.900000,0.996863416295580)(1.000000,0.999117761007151)
        };
        \addlegendentry{Unsquared}
    
    \end{axis}

\end{tikzpicture}
    }
    \caption{
    AUROC of CLAD as the margin hyperparameter is varied in the range \([0.1, 1]\), with (orange) and without (blue) the distance terms being squared.
    }
    \label{fig:hp_sweeps}
    \end{figure}
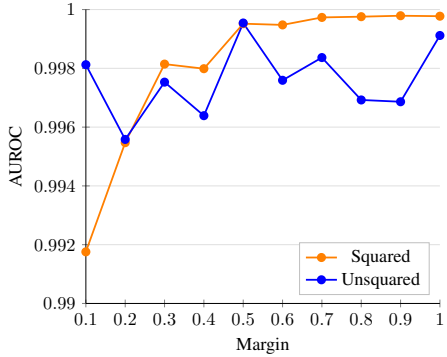

To empirically evaluate design choices used by the CLAD loss function, a margin based hinge regularisation term was introduced equivalent to the contrastive loss function. CLAD models were then trained with the margin value being swept in the range \([0.1, 1]\) in increments of \(0.1\), with and without the distance terms being squared. The results, shown in Figure~\ref{fig:hp_sweeps}, demonstrate that squaring the distance terms in the CLAD loss yields smoother and more stable improvements in AUROC across margins, whereas the unsquared formulation exhibits fluctuations in performance. Additionally, when the distance terms were squared, there was an increasing trend in the model’s performance as the margin value increases towards \(m = 1.0\). This suggests that the performance is greatest when the model is trained to maximally separate benign and malicious traffic such that the respective samples lie antipodal on the unit hypersphere. 

To evaluate the impact of tying the concentration terms such that \(\kappa_0 = \kappa_1\), a hyperparameter \(\alpha = \frac{\kappa_0}{\kappa_0 + \kappa_1}\) was defined, resulting in a loss function where positive-pair distances were scaled by \(\alpha\) and negative-pair distances were scaled by a factor of \(1-\alpha\). The \(\alpha\) parameter was then swept in the range \([0.1, 0.9]\). The mean AUROC across runs was found to be 0.999574 with a standard deviation of 0.000570 suggesting that CLAD is largely insensitive to the ratio of the concentration terms.

\subsection{Embedding Analysis}
\label{sec:exp_emb_analysis} 

\begin{figure}[!t]
\centering
\resizebox{\linewidth}{!}{
\begin{tabular}{cc} \footnotesize
\includegraphics[]{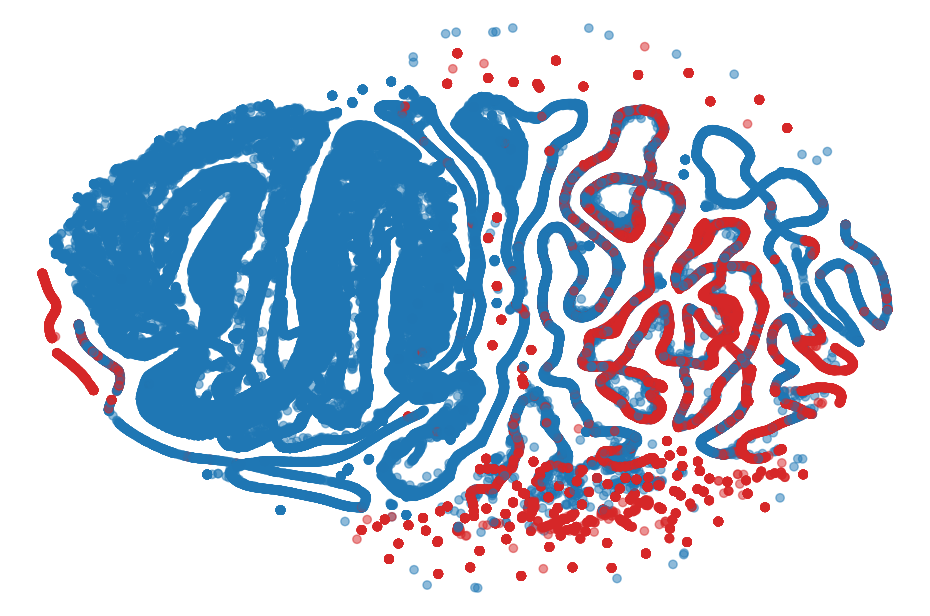} \hspace{1cm} &

\includegraphics[]{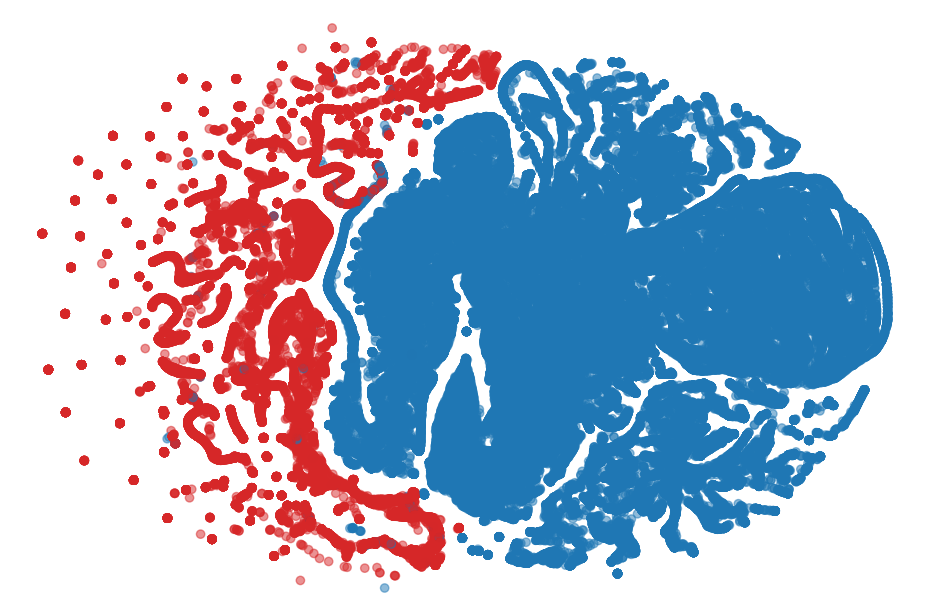}

\end{tabular}
}
\caption{T-SNE plots of the validation embeddings produced by \textbf{left:} the contrastive loss function and \textbf{right:} the CLAD loss function. Both loss functions produce distinct clusters for benign traffic~(blue) and malicious traffic~(red). The contrastive loss function embeddings have overlap between the clusters; while the CLAD loss function creates a clear separation between clusters.}
\label{fig:tsne_embedding_plot}
\end{figure}
A qualitative evaluation of the CLAD loss function was conducted by examining the output embeddings from the model's contrastive head. In Figure~\ref{fig:tsne_embedding_plot}, the embeddings generated by CLAD are visualized using t-SNE dimensionality reduction and compared against those produced using the contrastive loss. Visual inspection reveals that both loss functions are able to separate benign (blue) and malicious (red) traffic into distinct clusters. However, the contrastive loss exhibits some overlap between classes, whereas CLAD achieves a much clearer separation.

This distinction reflects the core advantage of the CLAD loss: by explicitly modelling benign traffic as a vMF distribution, the model is encouraged to structure the embedding space such that the centroid becomes a statistically meaningful and robust class proxy. Rather than relying on incidental spatial structure, CLAD imposes a probabilistic constraint that promotes both intra-class compactness and inter-class separation. While t-SNE offers qualitative insight, the improved detection performance and lower false positive rate quantitatively support the effectiveness of this representation.

\subsection{Alignment Analysis}
\label{sec:exp_alingnment_analysis}
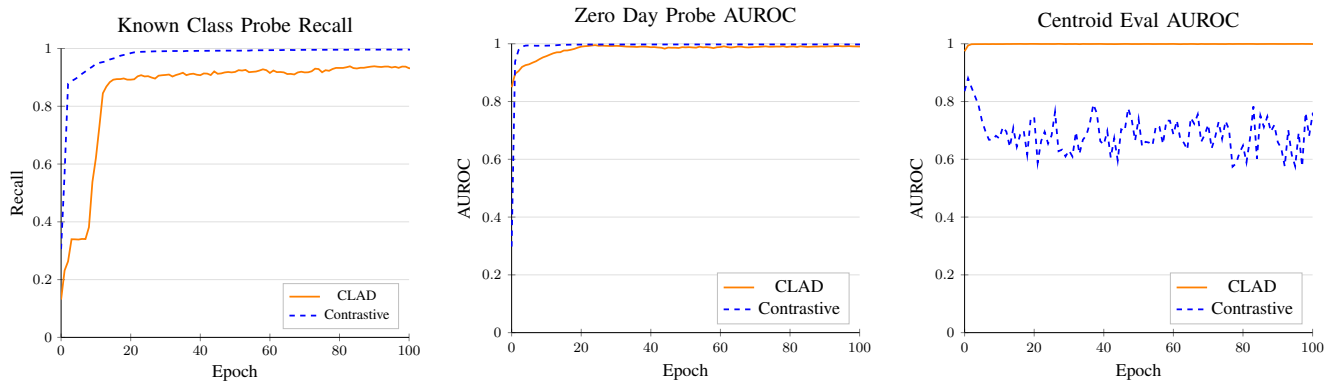
\begin{figure*}[!t]
\centering
\resizebox{\textwidth}{!}{
\begin{tabular}{ccc} \footnotesize
    \begin{tikzpicture}
    \begin{axis}[
        title = {Known Class Probe Recall},
        xlabel={Epoch},
        ylabel={Recall},
        xmin=0, xmax=100,
        ymin=0, ymax=1,
        title style={font=\fontsize{12}{14}\selectfont},
        xlabel style={font=\fontsize{10}{12}\selectfont},
        ylabel style={font=\fontsize{10}{12}\selectfont},
        x tick label style={font=\fontsize{8}{10}\selectfont},
        y tick label style={font=\fontsize{8}{10}\selectfont},
        legend pos=south east,
        axis lines=left, 
        axis line style={-},
        grid=none, 
        ymajorgrids=true, 
        major grid style={line width=.1pt, draw=gray!25}, 
        legend style={
            draw=gray!50, 
            line width=.1pt, 
        }
    ]

    \addplot[
        color=orange,
        mark=none,
        line width=1pt
        ]
        table [x=x, y=y, col sep=comma] {data/clad_alignment/clad_mal_recall.csv};
        \addlegendentry{CLAD}

     \addplot[
        color=blue,
        mark=none,
        dashed,
        line width=1pt,
        nodes near coords,
        point meta=explicit symbolic
        ]
        table [x=x, y=y, col sep=comma] {data/clad_alignment/contrastive_mal_recall.csv};
        \addlegendentry{Contrastive}

    \end{axis}
\end{tikzpicture}  &
    \begin{tikzpicture}
    \begin{axis}[
        title = {Zero Day Probe AUROC},
        xlabel={Epoch},
        ylabel={AUROC},
        xmin=0, xmax=100,
        ymin=0, ymax=1,
        title style={font=\fontsize{12}{14}\selectfont},
        xlabel style={font=\fontsize{10}{12}\selectfont},
        ylabel style={font=\fontsize{10}{12}\selectfont},
        x tick label style={font=\fontsize{8}{10}\selectfont},
        y tick label style={font=\fontsize{8}{10}\selectfont},
        legend pos=south east,
        axis lines=left, 
        axis line style={-},
        grid=none, 
        ymajorgrids=true, 
        major grid style={line width=.1pt, draw=gray!25}, 
        legend style={
            draw=gray!50, 
            line width=.1pt, 
        }
    ]

    \addplot[
        color=orange,
        mark=none,
        line width=1pt
        ]
        table [x=x, y=y, col sep=comma] {data/clad_alignment/clad_zd_auroc.csv};
        \addlegendentry{CLAD}

     \addplot[
        color=blue,
        mark=none,
        dashed,
        line width=1pt,
        nodes near coords,
        point meta=explicit symbolic
        ]
        table [x=x, y=y, col sep=comma] {data/clad_alignment/contrastive_zd_auroc.csv};
        \addlegendentry{Contrastive}

    \end{axis}
\end{tikzpicture}  &
    \begin{tikzpicture}
    \begin{axis}[
        title = {Centroid Eval AUROC},
        xlabel={Epoch},
        ylabel={AUROC},
        xmin=0, xmax=100,
        ymin=0, ymax=1,
        title style={font=\fontsize{12}{14}\selectfont},
        xlabel style={font=\fontsize{10}{12}\selectfont},
        ylabel style={font=\fontsize{10}{12}\selectfont},
        x tick label style={font=\fontsize{8}{10}\selectfont},
        y tick label style={font=\fontsize{8}{10}\selectfont},
        legend pos=south east,
        axis lines=left, 
        axis line style={-},
        grid=none, 
        ymajorgrids=true, 
        major grid style={line width=.1pt, draw=gray!25}, 
        legend style={
            draw=gray!50, 
            line width=.1pt, 
        }
        ]
    
    \addplot[
        color=orange,
        mark=none,
        line width=1pt
        ]
        table [x=x, y=y, col sep=comma] {data/clad_alignment/clad_centroid_auroc.csv};
        \addlegendentry{CLAD}

     \addplot[
        color=blue,
        mark=none,
        dashed,
        line width=1pt,
        nodes near coords,
        point meta=explicit symbolic
        ]
        table [x=x, y=y, col sep=comma] {data/clad_alignment/contrastive_centroid_auroc.csv};
        \addlegendentry{Contrastive}

    \end{axis}
\end{tikzpicture}  

\end{tabular}
}
\caption{Alignment analysis of the contrastive (blue) and CLAD (orange) loss functions when detecting known and zero-day attacks. 
\textbf{Left:} Balanced recall of the contrastive and CLAD loss functions amongst classes included in the training data when evaluated using a linear classification probe.
\textbf{Centre:} Mean AUROC of the contrastive and CLAD loss functions when evaluated on zero-day using a linear probe. 
\textbf{Right:} Mean AUROC of the contrastive and CLAD loss functions throughout training when evaluated by measuring the distance between validation samples and the centroid of benign traffic. 
}
\label{fig:alignment_results}
\end{figure*}

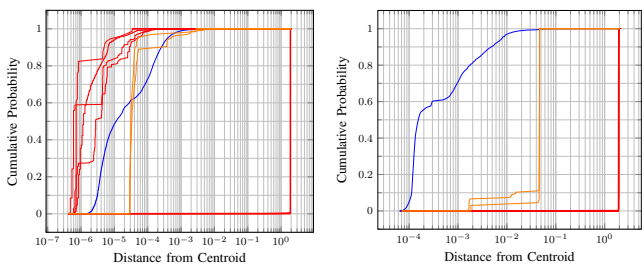
\begin{figure}[!t]
\centering
\resizebox{\linewidth}{!}{
\begin{tabular}{cc} \footnotesize
\resizebox{\linewidth}{!}{\begin{tikzpicture}
\begin{axis}[
    xlabel={Distance from Centroid},
    ylabel={Cumulative Probability},
    xlabel style={font=\fontsize{10}{12}\selectfont},
    ylabel style={font=\fontsize{10}{12}\selectfont},
    x tick label style={font=\fontsize{8}{10}\selectfont},
    y tick label style={font=\fontsize{8}{10}\selectfont},
    xmode=log, 
    log basis x=10,
    xticklabel={$10^{\pgfmathprintnumber{\tick}}$},
    grid=both, 
    minor tick num=1, 
    log ticks with fixed point, 
]

\addplot[
    color=blue,
    mark=none,
    line width=0.6pt,
] 
table [x=x, y=y, col sep=comma] 
{data/cdf_values/contrastive/class_0.csv};

\addplot[
    color=red,
    mark=none,
    line width= 0.6pt,
] 
table [x=x, y=y, col sep=comma] 
{data/cdf_values/contrastive/class_1.csv};

\addplot[
    color=red,
    mark=none,
    line width= 0.6pt,
] 
table [x=x, y=y, col sep=comma] 
{data/cdf_values/contrastive/class_2.csv};

\addplot[
    color=red,
    mark=none,
    line width= 0.6pt,
] 
table [x=x, y=y, col sep=comma] 
{data/cdf_values/contrastive/class_3.csv};

\addplot[
    color=red,
    mark=none,
    line width= 0.6pt,
] 
table [x=x, y=y, col sep=comma] 
{data/cdf_values/contrastive/class_4.csv};

\addplot[
    color=red,
    mark=none,
    line width= 0.6pt,
] 
table [x=x, y=y, col sep=comma] 
{data/cdf_values/contrastive/class_5.csv};

\addplot[
    color=red,
    mark=none,
    line width= 0.6pt,
] 
table [x=x, y=y, col sep=comma] 
{data/cdf_values/contrastive/class_6.csv};

\addplot[
    color=red,
    mark=none,
    line width= 0.6pt,
] 
table [x=x, y=y, col sep=comma] 
{data/cdf_values/contrastive/class_7.csv};

\addplot[
    color=red,
    mark=none,
    line width= 0.6pt,
] 
table [x=x, y=y, col sep=comma] 
{data/cdf_values/contrastive/class_8.csv};

\addplot[
    color=red,
    mark=none,
    line width= 0.6pt,
] 
table [x=x, y=y, col sep=comma] 
{data/cdf_values/contrastive/class_9.csv};
\addplot[
    color=orange,
    mark=none,
    line width= 0.6pt,
] 
table [x=x, y=y, col sep=comma] 
{data/cdf_values/contrastive/class_10.csv};

\addplot[
    color=orange,
    mark=none,
    line width= 0.6pt,
] 
table [x=x, y=y, col sep=comma] 
{data/cdf_values/contrastive/class_11.csv};

\end{axis}
\end{tikzpicture}} &
\resizebox{\linewidth}{!}{\begin{tikzpicture}
\begin{axis}[
    xlabel={Distance from Centroid},
    ylabel={Cumulative Probability},
    xlabel style={font=\fontsize{10}{12}\selectfont},
    ylabel style={font=\fontsize{10}{12}\selectfont},
    x tick label style={font=\fontsize{8}{10}\selectfont},
    y tick label style={font=\fontsize{8}{10}\selectfont},
    xmode=log, 
    log basis x=10,
    xticklabel={$10^{\pgfmathprintnumber{\tick}}$},
    grid=both, 
    minor tick num=1, 
    log ticks with fixed point, 
]

\addplot[
    color=blue,
    mark=none,
    line width=0.6pt,
] 
table [x=x, y=y, col sep=comma] 
{data/cdf_values/clad/class_0.csv};

\addplot[
    color=red,
    mark=none,
    line width= 0.6pt,
] 
table [x=x, y=y, col sep=comma] 
{data/cdf_values/clad/class_1.csv};

\addplot[
    color=red,
    mark=none,
    line width= 0.6pt,
] 
table [x=x, y=y, col sep=comma] 
{data/cdf_values/clad/class_2.csv};

\addplot[
    color=red,
    mark=none,
    line width= 0.6pt,
] 
table [x=x, y=y, col sep=comma] 
{data/cdf_values/clad/class_3.csv};

\addplot[
    color=red,
    mark=none,
    line width= 0.6pt,
] 
table [x=x, y=y, col sep=comma] 
{data/cdf_values/clad/class_4.csv};

\addplot[
    color=red,
    mark=none,
    line width= 0.6pt,
] 
table [x=x, y=y, col sep=comma] 
{data/cdf_values/clad/class_5.csv};

\addplot[
    color=red,
    mark=none,
    line width= 0.6pt,
] 
table [x=x, y=y, col sep=comma] 
{data/cdf_values/clad/class_6.csv};

\addplot[
    color=red,
    mark=none,
    line width= 0.6pt,
] 
table [x=x, y=y, col sep=comma] 
{data/cdf_values/clad/class_7.csv};

\addplot[
    color=red,
    mark=none,
    line width= 0.6pt,
] 
table [x=x, y=y, col sep=comma] 
{data/cdf_values/clad/class_8.csv};

\addplot[
    color=red,
    mark=none,
    line width= 0.6pt,
] 
table [x=x, y=y, col sep=comma] 
{data/cdf_values/clad/class_9.csv};
\addplot[
    color=orange,
    mark=none,
    line width= 0.6pt,
] 
table [x=x, y=y, col sep=comma] 
{data/cdf_values/clad/class_10.csv};

\addplot[
    color=orange,
    mark=none,
    line width= 0.6pt,
] 
table [x=x, y=y, col sep=comma] 
{data/cdf_values/clad/class_11.csv};

\end{axis}
\end{tikzpicture}} 
\end{tabular}
}
\caption{Cumulative distribution function of the cosine distance between data points and the centroid of benign traffic when training a model using the CLAD and contrastive loss functions. \textbf{Left:} The contrastive loss function learns a representation in which malicious classes (red) are often closer to the centroid than benign samples (blue). \textbf{Right:} the CLAD loss function learns an embedded representation of the data in which benign samples (blue) appear close to the centroid, malicious samples (red) are far from the centroid, and zero-day attacks (orange) are orthogonal to the centroid.}

\label{fig:alignment_box_plots}
\end{figure}

A quantitative comparison between the CLAD and contrastive loss functions was conducted by evaluating their respective abilities to classify known and zero-day network traffic throughout training. To facilitate this analysis, two linear classification probes were appended to the model alongside the contrastive projection head. One probe was trained on known attack classes in multiclass classification, whilst two additional malicious classes were separated from the training data and used to train the other probe as a binary classifier to identify zero-day attacks. Both probes were trained on embeddings after the gradient information had been detached, allowing them to evaluate the efficacy of the features learned by the model without impacting its weights. The performance of each probe was evaluated on the validation set. The performance of the contrastive head was measured by inferring class labels from the distance of validation samples to the centroid of benign embeddings in the training data, this method of evaluation is hereby referred to in the remainder of this work as centroid evaluation.

The results, given in Figure~\ref{fig:alignment_results}, indicate that optimising both the contrastive and the CLAD loss functions yields embedded representations where both known and zero-day classes are linearly separable. Notably, CLAD's representations of different malicious distributions are linearly separable, despite being learned using binary class labels. These findings were further validated by employing a KNN classifier to assess the performance of the contrastive heads, where CLAD and the contrastive loss were found to have comparable mean recall scores of \(.8877\) and \(.8769\) respectively.

While both models learned representations where benign and malicious traffic was linearly separable, only CLAD's contrastive head consistently improved in performance throughout training. In contrast, little correlation was observed between the performance of the contrastive loss function's centroid evaluation performance and its training progress, indicating that malicious embeddings often appear closer to the benign class centroid than benign embeddings. 

This phenomenon was further highlighted by plotting the cumulative distribution functions of the cosine distances from validation samples of each class to the benign centroid for both fully trained models, as shown in Figure~\ref{fig:alignment_box_plots}. The contrastive loss function learned an embedded representation in which many known malicious (mean \(= 1.157\), std \(= 0.926\)) and zero-day samples (mean \(= 2.458 \times 10^{-5}\), std \(= 4.415 \times 10^{-5}\)) lie closer to the benign centroid than a substantial fraction of benign traffic (mean \(= 0.00185\), std \(= 0.0587\)), explaining its poor performance when using centroid-based inference. Conversely, CLAD learned an embedded space in which benign traffic was tightly clustered around the centroid (mean \(0.00378\), std \(= 0.0634\)), with malicious traffic being maximally separated (mean \(= 1.922\), std \(= 0.0263\)). Zero-day attacks occupied the intermediate region (mean \(= 0.0431\), std \(= 0.0112\)), which is consistent with the orthogonal OOD score derived for CLOSR in Section~\ref{sec:proposed_multiclass}.

These findings demonstrate that while both the CLAD and contrastive loss functions generate discriminative features for identifying both known and zero-day attacks, the contrastive loss does not guarantee separation between malicious traffic and the benign class centroid. Hence, reference samples are required by the classifier to infer a zero-day attack's position in embedded space. This can be achieved via training a linear probe, or without re-training by using a KNN classifier. Conversely, the CLAD loss function ensures a high expectation of malicious samples appearing further from the benign centroid than benign samples, allowing zero-day attacks to be successfully detected. 

\subsection{Class Proxy Ablation}
\label{sec:additional_class_proxy}

\begin{table}[!t]
    \caption{AUROC comparison of CLAD using various class proxies.}
     
    \label{tab:class_proxy_ablation}
    \centering
    \begin{tabular}{cc}
    
    \hline 
    Class Proxy & Mean AUROC \\
    \hline
    
    Centroid & \textbf{.999633}\\
    Median & .999622\\
    Trimmed Mean & .999627 \\
    Medoid & .999627 \\
    Neighbour Distance & .645342 \\
    
    \hline
    \end{tabular}
    \end{table}
CLAD relies on the distance between a test sample and the centroid of benign traffic to perform inference, however, centroid-based classification methods can be sensitive to noisy training data and adversarial samples. In Table~\ref{tab:class_proxy_ablation}, the performance of CLAD is reported with alternative distance measures—specifically, the distance to the median, trimmed mean, medoid, and nearest neighbour. Amongst these, the centroid yields the best performance. 

Interestingly, the nearest neighbour distance performs worse than the centroid distance in this context, which contrasts with previous contrastive and self-supervised learning approaches. This can be attributed to CLAD explicitly modelling the benign traffic as a vMF distribution, unlike other non-parametric approaches such as the SupCon loss function. While these results suggest the centroid remains effective under clean conditions, future work will aim to evaluate the robustness of CLAD to adversarial samples and polluted benign samples in the training dataset.

\subsection{CLOSR Inference Ablation}
\label{sec:additional_closr_inference}
\begin{table}[!t]
    \caption{Open-set AUC of CLOSR using various OOD scoring functions.}
     
    \label{tab:ood_score_ablation}
    \centering
    \begin{tabular}{cc}
    
    \hline 
    OOD Score & Open-set AUC \\
    \hline
    
    Energy & .979926 \\
    Gaussian & .952326 \\
    Weighted Gaussian & \textbf{.993768} \\
    
    \hline
    \end{tabular}
    \end{table}

To identify zero-day attacks in the embedding space, CLOSR computes an OOD score by evaluating the likelihood of a test sample under a set of Gaussian distributions defined over the class-specific subspaces. This formulation is motivated by the Central Limit Theorem and the assumption that zero-day attacks occupy a broader region of the hypersphere relative to known classes. To empirically validate this assumption, two malicious classes were withheld from the training data to simulate zero-day attacks. A CLAD model was then trained, and the output embedding matrices were analysed. The known-class validation embeddings exhibited a normalised rank of 0.0625, indicating that known traffic concentrates within a small subspace of the hypersphere. In contrast, the zero-day embeddings yielded a substantially larger normalised rank of 0.4375, supporting the hypothesis that their distribution is significantly more diffuse.

To evaluate the suitability of the weighted Gaussian likelihood used by CLOSR as an OOD score, an ablation study compared this score against an energy-based score proportional to the negative joint log-likelihood of a test sample under all known class distributions~\cite{liu2021energybasedoutofdistributiondetection} and a score proportional to the likelihood of the sample under a Gaussian orthogonal to the centroid of the sample's predicted closed-set class label in its corresponding subspace. The resulting open-set AUC values are reported in Table~\ref{tab:ood_score_ablation}. The energy-based score outperformed the unweighted Gaussian score, suggesting that aggregating information across all class-specific subspaces provides more reliable zero-day detection. Finally, the soft-weighted Gaussian likelihood achieved the highest open-set AUC, indicating that a Gaussian-based formulation is a more appropriate OOD scoring mechanism than those derived from the vMF distributions used for modelling known classes.

\subsection{Computational Cost and Scalability}
\label{subsec:ablations_copmute}
\begin{table*}[!t]
    \caption{Computational requirements of CLAD, CLOSR, and baseline models during inference for binary classification (top) and open-set recognition (bottom).}
    \label{tab:compute_comp}
    \centering
    \resizebox{\linewidth}{!}{
    \begin{tabular}{cccccc}
    
    \hline 
    Model & Parameters (k) & VRAM (MiB) & Compute (MFLOPs) & Latency (ms) & Throughput (\(Ms^{-1}\)) \\
    \hline

    CLAD & 3318.80 & 33.06 & 6.61 & 0.39 & 2.31 \\
    Autoencoder & 5.91 & 9.91 & 0.01 & 0.20 & 80.32\\
    DUAD & 536.28 & 18.70 & 1.06 & 0.53 & 7.78\\ 
    Deep SVDD & 87.10 & 12.36 & 0.17 & 0.24 & 40.08 \\
    Renoir & 1181.28 & 31.01 & 7.03 & 2.10 & 2.25\\
    MLP & 667.16 & 19.08 & 1.33 & 0.20 & 10.06\\
    Siamese Network & 200.14 & 11.90 & 0.40 & 0.45 & 18.57\\
    \midrule
    CLOSR & 3048.20 & 33.18 & 6.07 & 0.40 & 2.68 \\
    DOC & 277.13 & 22.22 & 0.55 & 0.31 & 18.72\\
    OPENMAX & 655.24 & 20.92 & 1.31 & 2.57 & 10.94\\
    CROSR & 81.76 & 12.38 & 0.16 & 2.80 & 22.32\\
    Siamese Network & 3374.15 & 3392.09 & 109.82 & 19.06 & 2.29\\
    \hline
    \end{tabular}}
    \end{table*}

The computational requirements of CLAD, CLOSR, and baseline models during inference are summarised in Table~\ref{tab:compute_comp}, reporting parameter count, VRAM usage, computational complexity (MFLOPs), latency, and throughput. To ensure a fair comparison, all models were evaluated using the architectures obtained through hyperparameter optimisation. Measurement conditions were standardised: a fixed batch size of 1 was used for assessing computational complexity and latency, a batch size of 1024 was used for measuring VRAM usage, and throughput was determined using optimal batch sizes selected from \(\{2^i \mid i \in \mathbb{Z}, \ 1 \leq i \leq 20\}\). Latency and throughput were measured on an NVIDIA RTX 3090 GPU and Intel Xeon W-2255 CPU under Ubuntu 22.04, using PyTorch 2.0.1 (FP32). Each latency value is the mean over 20 timed runs after 5 warmups. FLOPs were estimated with the PyTorch profiler under identical model configurations.

In binary classification, the optimised CLAD architecture exhibited a substantially higher parameter count than the baseline models, reflecting its scalability to larger model sizes without overfitting. Despite this increased complexity, CLAD achieved lower latency and higher throughput than RENOIR, demonstrating clear efficiency advantages. Although it was less efficient than other baselines, its balance of low latency and comparatively high throughput highlights its suitability as a practical real-time NIDS solution.

In OSR, CLOSR was found to be less computationally efficient than non-contrastive baselines, which can partly be attributed to the hyperparameter search selecting a relatively large architecture. This reflects CLOSR’s robustness against overfitting. When compared to the Siamese network, CLOSR was markedly more efficient, achieving superior results across all computational metrics. The inefficiency of the Siamese network stems from its reliance on the neighbours-distance scoring function, which requires an expensive search over the training set. CLOSR's combination of latency and throughput makes it practical for NIDS. 

\section{Discussion and Conclusions}
\label{sec:Discussion and conclusion}

In this work the contrastive loss for anomaly detection~(CLAD) was introduced to bridge the gap between supervised classifiers, which perform well on known traffic but struggle to detect zero-day attacks, and anomaly detectors, which do not degrade in performance when detecting zero-day attacks but suffer from high false positive rates. The CLAD loss function exploits contrastive learning to exclusively model benign traffic as a von-Mises Fisher distribution in embedded space. As CLAD is only evaluated on benign anchor samples, the closed world assumption made by existing supervised classifiers is relaxed allowing for effective generalisation to zero-day attacks despite training on known malicious classes.

CLAD was experimentally evaluated and compared to leading anomaly detectors and supervised classifiers. It was found that by training on malicious samples, CLAD outperformed both existing supervised classifiers when detecting known attack classes. Furthermore, CLAD was able to effectively generalise to zero-day attacks, exploiting patterns learned from both known attacks and benign traffic to significantly outperform anomaly detectors when faced with previously unseen classes.

The extension of CLAD to open-set recognition gave rise to the Contrastive Loss for Open-Set Recognition~(CLOSR) framework, which enabled multiclass classification over a set known malicious classes whilst simultaneously identifying zero-day attacks. This was achieved by learning the distribution of each known class within a distinct embedded subspace, with classification performed by evaluating the likelihood of a test sample belonging to each distribution. Zero-day attacks manifested as embedded vectors orthogonal to the distributions' centroids allowing easy identification. Experimental results demonstrated that CLOSR not only maintains strong closed-set classification performance but also generalises effectively to zero-day attacks, thereby outperforming existing OSR approaches for network intrusion detection systems.

While the ability of CLAD and CLOSR to generalise to zero-day attacks represents a significant step toward practical machine learning–based network intrusion detection systems, several limitations remain. The primary drawback of CLOSR is its reliance on a distinct linear projection head for each class in the training data, leading to a model whose size and computational cost scales with the number of classes. This design choice may hinder cross-domain applicability, particularly in fields such as computer vision where datasets can contain thousands of classes. Moreover, the proposed approaches have not yet been evaluated against adversarial attacks or under conditions involving noisy training data, leaving their robustness in such scenarios uncertain. Addressing these gaps with future work could enhance both the efficiency and resilience of the methods. Finally, extending CLAD and CLOSR to support cross-network generalisation and few-shot learning on limited datasets would further strengthen their practicality and broaden their applicability.

\bibliographystyle{IEEEtran}
\bibliography{IEEEabrv, references}

\end{document}